\documentclass[twocolumn]{fairmeta}
\usepackage{amsmath}
\usepackage{pgfplots}
\usepackage{float} 
\usepackage{amssymb}
\usepackage{url}
\usepackage{enumitem}

\usepackage[math]{cellspace}

\DeclareMathOperator*{\argmax}{arg\,max}

\title{Memento: Personalized RAG-Style Long-Retention Data Scaling for Online Ads Recommendation}

\author[1]{Xiaoyu Chen}
\author[1]{Ruichen Wang}
\author[1]{Jieming Di}
\author[1]{Suofei Feng}
\author[1]{Nafis Abrar}
\author[1]{Lilly Kumari}
\author[1]{Tony Tsui}
\author[1]{Yilin Liu}
\author[1]{Yu Lu}
\author[1]{Sowmya Patapati}
\author[1]{Junwei Xiong}
\author[1]{Qiao Yang}
\author[1]{Dorothy Sun}
\author[1]{Yang Cao}
\author[1]{Victor Chen}
\author[1]{Pan Chen}
\author[1]{Ramsundar Sundarkumar}
\author[1]{Shivendra Pratap Singh}
\author[1]{Arnold Overwijk}
\author[1]{Ling Leng}
\author[1]{Dinesh Ramasamy}
\author[1]{Sri Reddy}
\author[1]{Robert Malkin}
\author[1]{Sandeep Pandey}

\affiliation[1]{Meta AI}

\abstract{
Modeling of long history data suffers from long-context window attention dilution, system efficiency and catastrophic forgetting problems, where naive linear scaling approach like LastN would fail. We introduce Memento, a personalized retrieval-augmented framework that treats historical user engagements as a document corpus and ad requests as queries, retrieving relevant interactions via Maximal Marginal Relevance (MMR) to balance similarity with diversity. We identify two complementary applications: \emph{Representation Memento}, which retrieves historical embeddings for feature augmentation, and \emph{Data Memento}, which retrieves past training examples for multipass training. Through infrastructure co-design---temporal chunking, INT8 quantization, and asynchronous serving---Memento achieves 5--10$\times$ resource efficiency over linear scaling. Memento processes daily requests with sub-10\,ms latency, yielding 0.25--0.3\% Normalized Entropy gain on both click-through and conversion prediction. In production, \textbf{Memento delivers a 1\% CTR lift on Facebook Feed and Reels and a 1.2\% CVR lift}, scaling personalization to 365+ days of history.
}

\date{\today}
\correspondence{First Author at \email{xochen@meta.com}}

\begin{document}

\maketitle

\section{Introduction}

Many companies have explored lifelong user history with prolonged data retention~\citep{pancha2022pinnerformer, chang2023twin, yongjiwu2021rethinking}, motivated by the intuition that a user's long-term behavioral patterns---seasonal shopping habits, evolving interests, life-stage transitions---contain signals that short-term windows miss entirely. The default strategy for data scaling is ``LastN''---retaining the last $N$ events or $N$ days. While simple and effective as an implicit noise filter, this recency-first approach tends to overfit~\citep{zhang2022oneepoch}, discards relevant historical context, and applies a uniform threshold to every user regardless of activity patterns. Generally speaking, there are three interrelated challenges:

\textbf{Needle in the haystack.} As the context window grows, relevant signals become increasingly sparse relative to noise. Transformer-based attention mechanisms~\citep{vaswani2017attention}, which underpin most modern sequence models, compute pairwise interactions across all positions in a sequence. When the sequence is long and most entries are irrelevant to the current prediction, the attention weights over the truly informative entries are diluted by the sheer volume of distractors. This degradation is not merely theoretical: studies show that even GPT-4 suffers significant retrieval accuracy loss when relevant information is embedded within long, irrelevant context ~\citep{arize2024needle}, and similar findings have been reported in academic research ~\citep{du2025context}. In our setting, a user's history may span hundreds of thousands of events, but only a handful are predictive for any given ad request---making this a particularly acute instance of the needle-in-the-haystack problem.

\textbf{Resource constraints and system efficiency.} Processing million-scale sequence lengths with prolonged data retention is computationally intractable for real-time ranking. Attention-based architectures incur $O(n^2)$ complexity in sequence length, and even with efficient variants~\citep{zaheer2020bigbird, beltagy2020longformer, dao2022flashattention}, the storage, compute, and serving of maintaining and processing full user histories grow linearly to quadratically with data volume. At scale, even small per-request overheads compound. A practical system must dramatically reduce the volume of history considered per request while preserving the most informative signals, analogous to how multi-stage retrieval and ranking pipelines handle candidate selection in search.

\textbf{Catastrophic forgetting.} Deep neural networks exhibit a well-documented tendency to lose previously learned knowledge when trained on new data~\citep{kemker2018measuring, kirkpatrick2017ewc}. This phenomenon, known as \emph{catastrophic forgetting}, is especially pronounced in recommendation systems where models are continuously trained on rolling time windows of data. As the training window advances, the model's parameters are updated to fit recent (user, ad, label) examples, and the gradient updates progressively overwrite representations of older behavioral patterns. Large recommendation models are often over-parameterized to achieve better in-distribution performance~\citep{belkin2019reconciling}, which paradoxically makes them more susceptible to forgetting: the excess capacity that enables tight fitting of current data also means that parameter regions encoding older knowledge are more readily repurposed. In practice, we have observed that models trained on a 30-day window can lose more than 50\% of their predictive performance on patterns from 6 months ago. Common mitigation strategies fall into two categories identified by~\cite{kemker2018measuring}: \emph{rehearsal}, which replays representative samples from earlier training data, and \emph{dual memory}, which maintains separate fast- and slow-learning components.

To address these three challenges, we introduce \textbf{Memento}: a personalized retrieval-augmented framework for long-retention data scaling in recommendation systems. The key insight behind Memento is that history scaling is fundamentally an information retrieval problem. Just as Retrieval-Augmented Generation (RAG) grounds LLMs with query-relevant external knowledge~\citep{lewis2020retrieval, guu2020realm}, Memento treats a user's full engagement history as a document corpus and personalized our context window of each ad request with highly-relevant history. This is achieved with Maximal Marginal Relevance (MMR)~\citep{carbonell1998mmr}, which balances similarity to the query with diversity among retrieved documents.

This retrieval-augmented paradigm yields two complementary applications. \emph{Representation Memento} (section \ref{sec:rep-memento}) retrieves relevant historical embeddings to augment the ranking model's feature set, providing a dual-memory solution to tackle forgetting issue. \emph{Data Memento} (section \ref{sec:data-memento}) retrieves forgotten training examples for second-pass rehearsal, directly addressing catastrophic forgetting with a rehearsal method to focus on relevant historical patterns. Both applications share the same underlying retrieval infrastructure and MMR formulation, but target different stages of the modeling pipeline---inference-time feature augmentation and training-time data augmentation, respectively.

We make the following contributions:
\begin{itemize}
    \item \textit{Algorithmic framing:} We formalize user history retrieval as a document retrieval problem, introducing a modified MMR objective that balances user-user similarity, user-ad similarity, and document diversity (Section~\ref{sec:design}).
    \item \textit{Dual application:} We demonstrate two complementary uses---\emph{Representation Memento} retrieves relevant historical embeddings for feature augmentation, while \emph{Data Memento} addresses catastrophic forgetting through retrieval-guided rehearsal (Section~\ref{sec:rep-memento} and Section~\ref{sec:data-memento}).
    \item \textit{Production system:} We describe the infrastructure co-design---including chunking, NormInt8 quantization, and asynchronous serving---that enables sub-10\,ms retrieval latency and 5--10$\times$ resource efficiency over linear scaling (Section~\ref{sec:production}).
    \item \textit{Empirical validation:} Memento achieves a 1\% CTR lift and a 1.2\% CVR lift, scaling personalization to 365+ days of history.
\end{itemize}

\section{Related Work}
\subsection{RAG in LLMs}
In the LLM domain, RAG~\citep{lewis2020retrieval, wu2022memorizing} has become the de facto method for incorporating external knowledge into language models. RAG decouples storage (an indexed corpus) from reasoning (the generative model), enabling dynamic knowledge injection without retraining. This paradigm has evolved into agentic LLMs that utilize tools, with traditional RAG representing the special case of tool-augmented systems using search engines for knowledge augmentation. Advances in RAG involve determining how to group similar chunks of information together and generating effective document and query encoders for retrieval.

\subsection{Lifelong User Modeling}

Recent work encodes lifelong user history into summarization models. TWIN~\citep{chang2023twin} and its successor~\citep{si2024twin} propose dedicated search layers for ultra-long context windows. These clustering-based methods assume queries and documents share the same embedding space, limiting flexibility and preventing fine-grained retrieval of specific historical events. Other approaches, such as PinnerFormer~\citep{pancha2022pinnerformer} and DV365~\citep{lyu2025dv365}, jointly train long-term user summarization with downstream tasks using temporal shifting to capture long-term interests. However, all these methods are affected by the forgetting problem: gradients from recent labels dominate, biasing representations toward recent patterns.

Our work draws intuition from both areas, with the following distinctions:
(1)~An explicit search module with a multi-purpose inverted index on lifelong user history, applicable to both user representation and training data augmentation;
(2)~Explicit modeling of query-to-document and document-to-document similarity through a modified MMR formulation;
(3)~Treating lifelong user history as augmentation, with explicit modeling of this additional context in the downstream ranking model.

\begin{figure*}[ht]
    \begin{center}
    \includegraphics[width=1.0\linewidth]{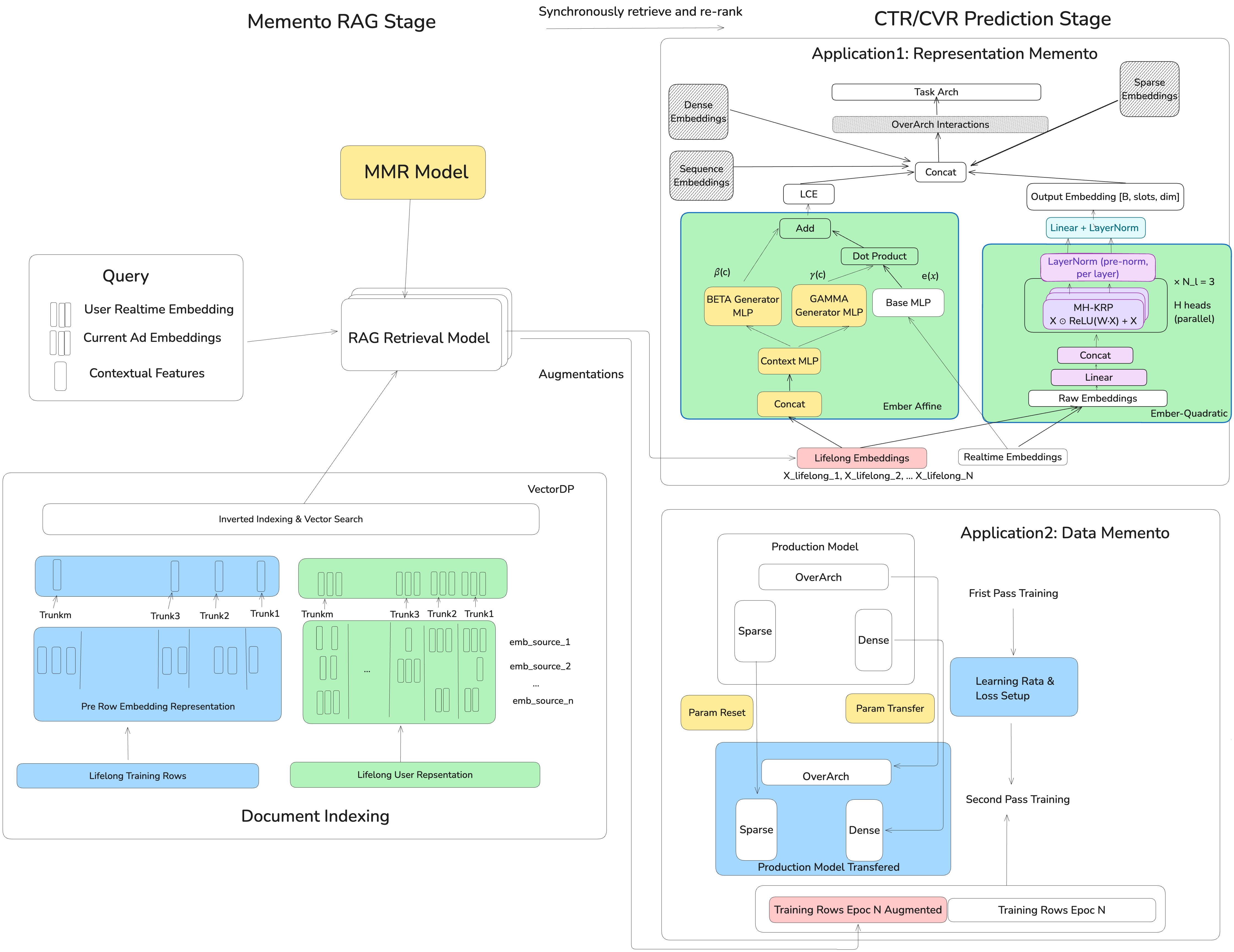}
    \caption{Memento detailed design diagram. \textbf{Left:Memento Retrieval Stage} Query patterns generate retrieval queries from ad requests; Memento indices organize user history into retrievable documents and retrieves through MMR; \textbf{Right: Application as Representation and Data Memento.} The application gives specific requirements through query and context, producing augmented user context for the ranking models in CTR/CVR prediction model}
    \label{fig:memento-design}
    \end{center}
\end{figure*}

\section{Memento Design}\label{sec:design}

The Memento platform adapts the core RAG paradigm---query formulation, document indexing, retrieval, and augmentation---to the recommendation setting (see Figure~\ref{fig:memento-design}). In traditional RAG for LLMs, a user question serves as the query, a text corpus serves as the document store, a retriever selects relevant passages, and the retrieved passages are prepended to the LLM's context for grounded generation. Memento mirrors this pipeline with four analogous components tailored to user history retrieval.

A \textbf{query} is generated from each ad request, encoding dimensions such as temporal recency, user cohort, ad category features, and interest topics. These multi-dimensional queries provide flexible retrieval options: for example, a query can emphasize topical relevance (``show me history related to this ad category'') or temporal recency (``show me the user's most recent engagement pattern''). The \textbf{Memento index} organizes a user's full engagement history into coherent, retrievable units across temporal and topical dimensions---analogous to document chunks in text-based RAG. Each index entry synthesizes user interactions over a defined time period and data source, serving as the atomic unit of retrieval. The \textbf{retrieval model} scores and selects the most relevant index entries for a given query, transforming sparse, noisy user event data into a compact set of high-quality signals. This component is extensible: it supports both applications of representation and data memento, as well as diverse similarity objectives across different user--entity relationships. Finally, the retrieved entries are transformed into \textbf{augmented user context}---embeddings, event sequences, or training data---that the downstream ranking model consumes as additional features or training examples.

In the remainder of this section, we describe the retrieval algorithms that power the retrieval model: MMR-based retrieval. We then present two applications of this framework: \emph{Representation Memento} (Section~\ref{sec:rep-memento}), which retrieves relevant user embeddings and events for inference-time feature augmentation, and \emph{Data Memento} (Section~\ref{sec:data-memento}), which retrieves historical training data for second-pass rehearsal against catastrophic forgetting. Both applications share the retrieval infrastructure but target different stages of the modeling pipeline.

\subsection{RAG-style Retrieval Model}
\subsubsection{MMR Model}

We represent documents and queries using intrinsic attributes, including traditional word tokens, pre-trained embeddings, and dense counter features. We then apply similarity functions to measure distances. We define three similarity functions:
\begin{equation}
 \text{Sim}_{UA}(Q, D), \quad
\text{Sim}_{UU}(Q, D), \quad
\text{Sim}_{UU}(D_i, D_j)
\end{equation}
These distinguish user-to-user similarity from user-to-ad similarity. The query can use only user-side information (e.g., current user interests) or incorporate ad-side information (e.g., target ad taxonomy), which significantly affects retrieval efficiency for different applications. We use a modified MMR scoring function that iteratively scores and retrieves documents, balancing diversity and similarity:

\begin{multline}\label{eq:mmr}
    \text{MMR}  = \argmax_{D_i \in R \setminus S} \left[ \alpha \cdot \text{Sim}_{UU}(D_i, Q) \right. \\
     + \beta \cdot \text{Sim}_{UA}(D_i, Q) \\
    \left. - (1 - \alpha - \beta) \cdot \max_{D_j \in S} \text{Sim}_{UU}(D_i, D_j) \right]
\end{multline}

where $Q$ represents the query, $D$ is the document (user history item), $S$ is the set of already selected items, $R$ is the candidate set, and $\text{Sim}$ denotes cosine similarity between embeddings. The hyperparameters $\alpha$ and $\beta$ control the trade-off: $\alpha$ weights user-to-user similarity, $\beta$ weights user-to-ad similarity, and $(1 - \alpha - \beta)$ weights diversity.

We integrate user document history into VectorDP, Meta's internal vector database platform, enabling inverted index construction and sub-10\,ms vector search and retrieval at scale.

We have implemented our unsupervised MMR in two configurations:
\begin{enumerate}
    \item \textbf{In-model MMR}: Used for parameter tuning and exploration during the experimental phase.
    \item \textbf{Infra MMR}: A precomputed solution for production to minimize training and inference latency.
\end{enumerate}

\subsubsection{Retrieval Infra Codesign}
The personalized retrieval system is built upon the co-design and integration of three specialized infrastructure components:
\begin{itemize}[noitemsep]
    \item \textbf{Data Generation and Aggregation:} A unified event aggregation infrastructure processes both batch and streaming user events to create "mementos," which are coherent, synthesized representations of user history (e.g., user embeddings over specific time windows). This infrastructure ensures high data freshness, with event processing and selection occurring with high frequency (e.g., 12-hour freshness). It also implements techniques like Maximal Marginal Relevance (MMR) for diversity-aware selection, ensuring the retrieved context is both relevant to the current request and non-redundant.
    \item \textbf{Vector Indexing and Storage:} A dedicated vector data platform is employed to build and maintain efficient indices for similarity search over the massive user history corpus. Key features include distributed index building, support for flexible clustering and sharding (e.g., K-Means), and a two-level indexing architecture (Gateway and Leaf layers) to support ultra-low latency, trillion-scale queries. This component is optimized for efficiency through methods like data compression (e.g., 8-bit integer quantization) while supporting real-time index updates.
    \item \textbf{Data Serving and Query Engine:} A unified generator serving framework is used for real-time and multi-hop data fetching. This system includes a flexible query layer that orchestrates complex online computation and retrieval across various backend services (e.g., unified data stores, vector indices). It supports a wide range of data operators (e.g., KNN similarity search, graph random walk) and utilizes a ranking layer to select the highest-quality ad candidates from multiple generators based on a latency budget.
\end{itemize}

\subsection{Application 1: Representation Memento}\label{sec:rep-memento}

\subsubsection{Retrieval Stage}

User histories are summarized into embeddings using an architecture similar to~\cite{zhang2024scaling}. These embeddings are generated on a daily schedule with dimension sizes typically ranging from 64 to 256, encoding user interests for each day. Different sources are distinguished by objective task and traffic type, numbering approximately 40. With 365 days of history, the effective sequence length reaches $O(100\text{K})$. These embeddings are integrated into VectorDP for efficient KNN search through MMR.

We apply \emph{chunking}---temporal aggregation of consecutive days into coarser epochs---to reduce retrieval granularity. Through our analysis of average embedding similarity across days and sources, we adopt a 7-day time-epoch-based chunking strategy that balances compression with information retention.

MMR is conducted by computing cosine similarity between the realtime user embedding and historical embeddings, as well as between the realtime ad embedding and historical embeddings. With MMR, we reduce embeddings by 50\% without model performance loss. Combined with chunking and NormInt8 quantization, this yields a significant improvement over linear scaling.



\subsubsection{Ember Architecture}

Beyond directly concatenating lifelong embeddings with the original features, we introduce a two-stage projection architecture named \textbf{Ember} (Memento Blended Enhanced Representations) to efficiently leverage retrieved embeddings while addressing computational complexity.

\textbf{Ember---Affine}: Context-aware Feature Modulation using affine transformation~\citep{perez2018film}. In production ads ranking models, embeddings are usually just concatenated with the rest of the user-side features and projected independently — there's no mechanism for the lifelong signal to actually shape how other features are represented.
Simple concatenation leaves that capacity on the table, while standard interaction layers (e.g., DCN, cross-attention) are too expensive to run at the embedding layer in a model that already serves billions of requests per day.
Ember---Affine sidesteps this cost by letting the lifelong context $c$ \emph{modulate} each user embedding $e(x)$ through a simple affine transform, rather than interacting with it through attention or crossing. A small MLP conditioned on $c$ produces per-dimension scale ($\gamma$) and shift ($\beta$) parameters, which are applied elementwise:                                                                                
    \begin{equation}                                      
        e'(x; c) = \gamma(c) \odot e(x) + \beta(c)
    \end{equation}

The MLP itself is deep with RMSNorm and residual connections for training stability, and is initialized so that $\gamma{=}1$ and $\beta{=}0$ --- the modulated embedding $e'(x; c)$ starts as an identity pass-through of $e(x)$ and the model learns when to deviate. 
Each modulated feature has its own MLP head producing $(\gamma_i, \beta_i)$ from the shared context $c$, and the modulation itself is element-wise --- keeping the per-feature cost at $O(d)$ and avoiding the pairwise interactions that make attention or feature crossing impractical at the embedding layer.   



\textbf{Ember---Quadratic}: Explicit cross-feature interactions via Quadratic Neural Networks (QNN)~\citep{li2025qnn}. In our current Embedding Feature Architecture (EFA), explicit cross-feature interaction only happens \emph{after} aggressive compression. The interaction stack --- LCE compression, followed by dot-product compress(DCPP) $X(X^{\top} Y + Z)$ whose weight matrices $Y, Z$ are generated by MLPs over the compressed input, feeding a deep ResidualMLP --- has no explicit feature interaction before compression. By the time DCPP runs, the $\sim\!3{,}800$ input embedding slots have been linearly compressed to $64$ slots (a $\sim\!60{:}1$ ratio), so signal that depends on the uncompressed structure of individual features is lost before interaction. We close this gap with a Quadratic Neural Network (QNN) module~\citep{li2025qnn} that runs in parallel to LCE on a curated set of raw user embeddings to compute explicit cross-feature interactions before compression. 
  

The final interaction equation of Ember---Quadratic adopts the T19 quadratic neuron from \citet{li2025qnn}: a Hadamard-product quadratic with mid-activation and an additive skip. Each layer applies a ReLU-gated quadratic transform, instantiated per head to reduce the interaction cost from $\mathcal{O}(d K^{2})$ to $\mathcal{O}(d K^{2} / H)$:
\begin{equation}
  \Psi(X^{h}) \;=\; X^{h} \odot \mathrm{ReLU}\!\left(W_{a}^{h} X\right) + X^{h}.
\end{equation}

The output is consumed by the DCPP's weight-generation path: which lets the DCPP derive its weights from \emph{both} the LCE-compressed view and the QNN cross-feature view, while preserving the $X X^{\top}$ outer product as an independent learning signal. Production deployment parameters (selected feature count, projection dimension, layer and head counts, integration slot budget) are reported in Table~\ref{tab:ember-deployment}.

\subsection{Application 2: Data Memento}\label{sec:data-memento}

As discussed in the Introduction, recommendation models trained on rolling time windows suffer from catastrophic forgetting: the model progressively loses knowledge of earlier behavioral patterns as parameter updates overwrite representations learned from older data. In our production models, this manifests as measurable degradation in predictive performance on user cohorts whose behavior is best explained by historical patterns (e.g., seasonal shoppers, users with long purchase cycles). The standard mitigation---simply extending the training window---is limited by the computational cost of training on larger datasets and, more fundamentally, by the fact that uniform data scaling does not prioritize the most informative historical examples.

Data Memento addresses this by treating the forgetting problem as a retrieval problem. Rather than training on all historical data indiscriminately, we selectively retrieve the most relevant past training examples and replay them as augmentation during second-pass training---a targeted form of the \emph{rehearsal} strategy identified in the continual learning literature~\citep{kemker2018measuring, lopezpaz2017gem}. Each training row consists of a (user, ad, label) tuple, and the model trains from older timestamps toward the latest timestamp sequentially. During second-pass training, Memento retrieves ``forgotten'' historical examples that are most relevant to the current data distribution, re-exposing the model to patterns it would otherwise lose. We demonstrate that this approach substantially improves model generalization (evaluation NE) compared to both uniform random replay and no replay at all.

\subsubsection{Retrieval Stage}
We first represent training rows as dense embeddings. We train a standard two-tower model to produce user embeddings and ad embeddings as representations of each row. We further chunk training rows by hour, matching the natural partitions of our offline tables. During chunking, we apply positive and negative sampling of rows based on model loss, inspired by prior work~\citep{paul2021deep} that quantifies the forgetting effect through model loss. We then index training data embeddings in VectorDP for inverted index construction and fast KNN search.

MMR scores are calculated by measuring cosine similarity between recent and historical training row embeddings, from both the user side and the ad side. During offline training, augmented rows are retrieved and provided to the model as additional training examples.

\subsubsection{Second-Pass Training with Knowledge Replay}
The second-pass training procedure is designed to integrate retrieved historical examples without allowing the model to overfit to them. Following~\cite{fan2024multi}, we reset the deep neural network's sparse parameters (embedding tables) before the second pass. This reset serves a dual purpose: it prevents the model from simply memorizing the replayed examples through its large embedding capacity, and it forces the dense layers to learn generalizable patterns from the combined recent and historical data. We also evaluated an alternative approach---embedding shrinking, which multiplies embedding parameters by a small fraction (e.g., 0.1) rather than fully resetting them---but as shown in our experiments (Section~\ref{sec:eval}), full resetting outperforms shrinking by 0.04\% NE, suggesting that the residual information in shrunken embeddings introduces more overfitting risk than useful signal.

During each recurring training epoch, we first apply parameter resetting with an enlarged learning rate to accelerate re-learning, then incorporate MMR-retrieved historical examples alongside recent data for the second training pass. The enlarged learning rate compensates for the information lost during resetting, while the retrieval-selected historical examples provide the model with targeted rehearsal on the patterns most at risk of being forgotten.

\section{Evaluation}\label{sec:eval}
\subsection{Experimental Setup}
\textbf{Dataset}. All experiments were conducted on industrial datasets with online advertising data. Each training row is a triplet of (user, ad, label), where label $= 1$ for the CTR task if an impression is clicked, and label $= 1$ for the CVR task if a click leads to a conversion. We did not use public datasets due to incompatibilities with our serving system and substantial discrepancies with internal model architectures.

All models are trained on approximately 50B examples for full convergence, with evaluation performed on the next billion-level examples. For Representation Memento, documents consist of user history with an effective sequence length of $O(100\text{K})$, comprising approximately 20K embedding sequences and approximately 190K raw event sequences. For Data Memento, documents span 365 days of training data at exabyte scale.

\textbf{Evaluation Metrics: Normalized Entropy}
We use Normalized Entropy (NE) as the evaluation metric for offline model performance~\citep{he2014practical, zhang2022dhen}. NE measures the ratio of the model's average log-loss per impression to the log-loss of a constant model predicting the background CTR. Lower NE indicates better predictive performance.
\begin{equation}
\text{NE} = \frac{-\frac{1}{N}\sum_{i=1}^{N} \left[ y_i \log(p_i) + (1-y_i)\log(1-p_i) \right]}{-\left[ p\log(p) + (1-p)\log(1-p) \right]}
\end{equation}
where $N$ is the total number of examples, $y_i \in \{0, 1\}$ are the labels for $i = 1, 2, \ldots, N$, $p_i$ is the estimated click probability for each impression, and $p$ is the average empirical CTR (or CVR, depending on the task).
\begin{figure}[ht]
\centering
\begin{tikzpicture}
    \begin{axis}[
        title={CTR Model Retention Scaling},
        xlabel={Retention (days)},
        ylabel={NE\%},
        xmin=0, xmax=365,
        ymin=-0.4, ymax=0.1,
        grid=major,
        every mark/.append style={fill=blue},
        legend pos=north west,
        width=7cm
    ]
    \addplot[
        color=blue,
        mark=*,
        smooth
    ]
    coordinates {
        (30, 0)
        (50, -0.06)
        (70, -0.1)
        (180, -0.11)
        (365, -0.11)
    };
    \addlegendentry{LastN Linear Scaling}

     \addplot[
        color=red,
        mark=*,
        smooth
    ]
    coordinates {
        (30, 0)
        (50, -0.1)
        (90, -0.13)
        (180, -0.2)
        (365, -0.25)
    };
    \addlegendentry{RAG-Style Scaling}
    \end{axis}
    \begin{axis}[
        title={CVR Model Retention Scaling},
        xlabel={Retention (days)},
        ylabel={NE\%},
        xmin=0, xmax=365,
        ymin=-0.4, ymax=0.1,
        grid=major,
        every mark/.append style={fill=blue},
        legend pos=north west,
        width=7cm,
        at={(0,7cm)}
    ]
    \addplot[
        color=blue,
        mark=*,
        smooth
    ]
    coordinates {
        (40, 0)
        (90, -0.14)
        (180, -0.196)
        (365, -0.2)
    };
    \addlegendentry{LastN Linear Scaling}

     \addplot[
        color=red,
        mark=*,
        smooth
    ]
    coordinates {
        (40, 0)
        (90, -0.2)
        (180, -0.25)
        (365, -0.28)
    };
    \addlegendentry{RAG-Style Scaling}
    \end{axis}
\end{tikzpicture}
\captionof{figure}{Retention scaling comparison: LastN linear scaling vs.\ RAG-style scaling for CTR (bottom) and CVR (top) models. Lower NE\% indicates better performance relative to baseline.}
\label{fig:retention-scaling}
\end{figure}
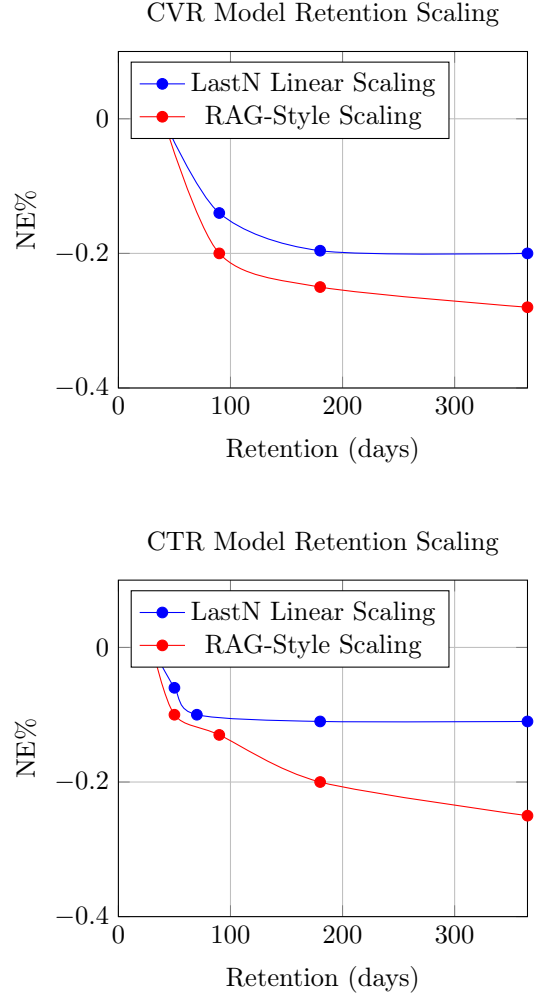

\subsection{RAG Performance}
We first compare RAG-style scaling against LastN linear scaling. Our baseline CTR model trains on 30 days of retention, while the CVR model trains on 40 days. For the CTR model, LastN linear scaling saturates at approximately 70-day retention, beyond which further scaling yields minimal gain. For the CVR model, LastN scaling saturates at 180 days---much later than the CTR model due to the sparsity of click-conditioned conversion data. As shown in Figure~\ref{fig:retention-scaling}, RAG-style scaling outperforms LastN by a substantial margin for both models, achieving 0.25--0.30\% NE gain at full 365-day retention.

Next, we present the performance fine-tuning of the RAG retrieval model. The RAG configuration for MMR is specified in the format MMR@filter-rate, \{$\alpha$, $\beta$\}, where a filter rate of 0.5 retains 50\% of documents. Our baseline is MMR@0.5 \{0, 0\}, which filters with maximum cross-document diversity. This baseline alone demonstrates the effectiveness of our unsupervised MMR method in addressing the needle-in-the-haystack problem.

As shown in Table~\ref{table:rag_roi}, at a filter rate of 50\%, combining user-to-user similarity $\text{Sim}_{UU}(D_i, Q)$ with user-to-ad similarity $\text{Sim}_{UA}(D_i, Q)$ slightly outperforms using either alone (e.g., \{0, 0.4\} vs.\ \{0, 0\} at MMR@0.5). When the filter rate increases to MMR@0.25, the benefit of combining both similarity functions becomes more pronounced, with $>$0.06\% NE improvement comparing \{0.1, 0.8\} to \{0, 0\}. However, if $\beta$ is too high (e.g., 0.95), MMR performance degrades below the diversity-only baseline.

\begin{table}[h]
    \centering
    \begin{tabular}{Sc|ScSc}
        \hline

        \multicolumn{3}{Sc}{\textbf{MMR@0.5 (50\% retrieval)}} \\
        \hline
        \{$\alpha$, $\beta$\}  & CTR N.E. & CVR N.E. \\
        \hline
        \{0, 0\} (baseline) & -- & -- \\
        \{0.3, 0\} & +0.025\% & +0.021\% \\
        \{0.7, 0\} & +0.03\% & +0.03\% \\
        \{0, 0.4\} & \textbf{-0.0\%} & \textbf{-0.01\%} \\
        \hline
        \multicolumn{3}{Sc}{\textbf{MMR@0.25 (25\% retrieval)}} \\
        \hline
        \{$\alpha$, $\beta$\}  & CTR N.E. & CVR N.E. \\
        \hline
        \{0, 0\} (baseline) & -- & -- \\
        \{0.1, 0.5\} & -0.02\% & -0.022\% \\
        \{0.05, 0.8\} & \textbf{-0.06\%} & \textbf{-0.067\%} \\
        \{0.0, 0.95\} & +0.01\% & +0.025\% \\
        \hline
    \end{tabular}
    \caption{RAG fine-tuning with different filter rates and parameter configurations. NE values are relative to the diversity-only baseline MMR@0.5 \{0, 0\}; negative values indicate improvement (lower NE is better). Positive values indicate degradation.}
    \label{table:rag_roi}
\end{table}

\subsection{Representation Memento Performance}
For Representation Memento, we conduct an extensive study on the effects of components beyond RAG configuration: embedding source scaling, Ember architecture variants, and efficiency optimizations (chunking, quantization). In addition to NE, we report infrastructure metrics including model inference QPS and storage measured in number of floats per user. All configurations apply MMR RAG by default.

As shown in Table~\ref{tab:embeddingmemento}, we evaluate several versions: LITE is a minimum viable version scaling only 8 embedding sources. V1 retrieves 60 embeddings via MMR@0.5. V1-TQ adds 7-day chunking and NormInt8 quantization, which is approximately 50\% more efficient than FP16. Ember architecture variants (Quadratic, Affine) are tested on top of V1-TQ.

NE gain scales with embedding count from 8 to 60. chunking and quantization reduce storage from 701K to 50K floats per user---a 14$\times$ reduction---while preserving model quality. Ember architectures provide an additional 0.03\% NE gain with improved inference QPS.

\begin{table*}[ht]
    \centering
    \begin{tabular}{Sc|ScScScSc}
        \hline
        \multicolumn{5}{Sc}{Experiment Results}\\
        \hline
         Name & CTR N.E. & CVR N.E. & Inference QPS & Storage \\
        \hline
         LITE & -0.05\% & -0.08\% & -0.5\% & 93K \\
         V1 & -0.2\% & -0.17\% & -2.28\% & 701K \\
         V1-TQ & -0.19\% & -0.168\% & -2.28\% & 50K \\
         V1-TQ-Ember(Affine) & -0.22\% & -0.23\% & -5.28\% & 50K\\
         V1-TQ-Ember(Quadratic) & -0.22\% &  -0.26\% & -5.28\% & 50K\\
         V1-TQ-Ember(Affine + Quadratic) & \textbf{-0.25\%} & \textbf{-0.26\%}  & -7.00\% & 50K \\
        \hline
    \end{tabular}
    \caption{Representation Memento experiments with different configurations. NE values are relative to the baseline (no Memento); negative indicates improvement. Inference QPS shows percentage change from baseline (negative indicates fewer queries served). Storage is in number of float-equivalent values per user.}
    \label{tab:embeddingmemento}
\end{table*}

\subsection{Data Memento Performance}
Data Memento is evaluated primarily on CVR models due to the scale of training data involved. As described earlier, we apply a multipass setup for the CVR model, where additional training data is augmented during second-pass training to improve generalization. We study the effects of multiple variants, including second-pass data scaling and model resetting strategies.

Our baseline CVR model is the production-deployed model trained on more than 150 days of data, fully saturated. The MP (Multipass) baseline resets the production model's sparse parameters and retrains on the most recent 35 days. DM (Data Memento) variants further augment this process with an additional 25\% of historical training data: RAND25 selects data randomly, while MMR25 selects via the MMR algorithm from $O(365)$ days of history. We compare resetting (RS) against embedding shrinking (ES), which multiplies embeddings by 0.1 to reduce the norm.

DM-MMR25-RS achieves the best result with 0.195\% NE gain. Embedding shrinking exhibits high run variance, so numbers are averaged across three runs. Resetting outperforms shrinking by 0.04\% NE; MMR outperforms random selection by 0.07\% NE; and 25\% data augmentation in the second pass nearly doubles the gain compared to the multipass baseline alone.

\begin{table}[H]
    \centering
    \begin{tabular}{Sc|Sc}
        \hline
        Setup  & Eval N.E. \\
        \hline
        MP & -0.107\%  \\
        DM-RAND25-RS & -0.12\%  \\
        DM-MMR25-ES  & -0.155\% \\
        DM-MMR25-RS &  \textbf{-0.195\%} \\
        \hline
    \end{tabular}
    \caption{Data Memento experiments for the CVR model. NE values are relative to the V0 baseline (150+ day trained model); negative indicates improvement. MP = Multipass baseline (reset + 35-day retrain). DM variants add 25\% historical data: RAND = random sampling, MMR = MMR-based selection. RS = parameter reset, ES = embedding shrinking ($0.1\times$ norm). Results averaged over 3 runs.}
    \label{tab:datamemento}
\end{table}

\section{Production Deployment and Online Performance}\label{sec:production}
\textbf{History Data Preparation.}
To support scalable aggregation, truncation, and update of user history, the platform accesses historical user events and embeddings through a highly scalable key-value abstraction with flexible triggering mechanisms. User embedding sequences are truncated to a bounded representation that balances historical coverage and retrieval efficiency, preventing unbounded growth while preserving the most informative signals for long-term retrieval. Updates are scheduled incrementally, allowing history representations to evolve continuously without blocking online serving.

To make heterogeneous histories usable for retrieval, we remove duplicate update triggers across sources and align updates to a uniform temporal interval. This alignment simplifies downstream retrieval and avoids redundant representation updates. The resulting embeddings are normalized into a row-wise representation, enabling consistent indexing and retrieval across sources.

\textbf{Online Data Serving and Retrieval.}
Indices are generated asynchronously and rebuilt when history updates are applied. Because index construction is decoupled from the online serving path, rebuilds do not block inference or retrieval. At serving time, retrieval operates on the latest available index snapshot, ensuring stable and predictable behavior.

During online inference, Memento performs nearest-neighbor retrieval over indexed history embeddings. Auxiliary operations such as query construction are executed asynchronously and in parallel with model computation. As a result, retrieval completes within a 10\,ms budget, introducing negligible end-to-end serving overhead.

\textbf{Online Performance.}
For Representation Memento, we prepared two deployment versions: V1-LITE, a single embedding source with $O(180)$-day retention for smaller models with limited storage and serving budgets; and V1-FULL, the full set of 30+ embedding sources with $O(360)$-day retention and Ember architecture for top-tier models. Across all production deployments, we observe no statistically significant ($<$5\%) inference QPS increase.

For Data Memento, deployment is limited to CVR models due to the large compute and storage for CTR models. We have deployed $O(360)$+ day retention for second-pass data scaling. Aggressive down-sampling has been applied based on MMR (25--50\%) so that training latency increase remains within budget and can be mitigated with additional trainers.

Through several rounds of large-scale A/B testing, Memento has been shipped to our top production CTR and CVR models, demonstrating a 1\% CTR lift and a 1.2\% CVR lift on Offsite Conversion. These results validate that RAG-style long-retention data scaling is effective at production scale.

\section{Conclusion}
In this paper, we presented Memento, a personalized retrieval-augmented framework for long-retention data scaling in online advertising. By reframing history scaling as an information retrieval problem, Memento selectively retrieves relevant historical signals using MMR. We demonstrated effectiveness through two complementary applications: \emph{Representation Memento} and \emph{Data Memento}. Both applications demonstrate that retrieval-based history scaling continues to improve where linear last-$N$ retention saturates.

 Memento processes daily requests with sub-10\,ms latency, delivering a 1\% CTR lift and 1.2\% CVR lift in production. Looking ahead, we aim to improve the retrieval algorithm through supervised approaches and extend the framework to additional data sources and platforms.

\clearpage
\newpage
\nocite{*}
\bibliographystyle{assets/plainnat}
\bibliography{paper}
\clearpage
\newpage

\end{document}